\documentclass[structabstract]{aa}  
\usepackage{graphicx}
\usepackage{txfonts}
\usepackage{natbib}
\bibpunct{(}{)}{;}{a}{}{,} 
\begin{document}
\title{Mixing of CNO-cycled matter in massive stars\thanks{Based on
observations obtained at the European Southern Obser\-vatory, proposals
62.H-0176 \& 074.B-0455(A).}\fnmsep\thanks{Based on observations collected at the Centro
Astron\'omico\,Hispa\-no 
Alem\'an at Calar Alto, proposals H2001-2.2-011 \& H2005-2.2-016.}} 


\author{N. Przybilla\inst{1}
\and
M. Firnstein\inst{1}
\and
M.F. Nieva\inst{2}
\and
G. Meynet\inst{3}
\and
A. Maeder\inst{3}
       }

\offprints{przybilla@sternwarte.uni-erlangen.de}

\institute{Dr. Karl Remeis-Observatory \& ECAP, Astronomical Institute, 
	   Friedrich-Alexander University Erlangen-Nuremberg,
           Sternwart\-str.~7, D-96049 Bamberg, Germany
\and
           Max-Planck-Institut f\"ur Astrophysik,
           Karl-Schwarzschild-Str. 1, D-85741 Garching, Germany
\and
           Geneva Observatory, University of Geneva, 
	   Maillettes 51, CH-1290 Sauverny, Switzerland
          }

\date{Received ; accepted }

 
  \abstract
  {}
   {We test predictions of evolution models on mixing of
    CNO-cycled products in massive stars from a fundamental perspective.
    Relative changes within the theoretical C:N:O abundance ratios and the buildup of helium
    are compared with observational results.}
   {A sample of well-studied Galactic massive stars is presented.
    High-quality optical spectra are carefully analysed 
    using improved NLTE line-formation and comprehensive analysis strategies.
    The results are put in the context of the existing literature data.}
   {A tight trend in the observed $N/C$ vs. $N/O$ ratios and the buildup of helium
    is found from the self-consistent analysis of main-sequence to supergiant
    stars for the first time. The catalytic nature of the CNO-cycles
    is confirmed quantitatively, though further investigations are
    required to derive a fully consistent picture.
    Our observational results support the case of strong mixing, as predicted 
    e.g. by evolution models that consider magnetic fields or by models that have 
    gone through the first dredge-up in the case of many supergiants.}
    {}


   \maketitle
%

\section{Introduction}
Energy production in massive stars is governed by the CNO-cycles
throughout most of their lifetimes.
The general correctness of our understanding of the CNO cycles and
of the relevant nuclear data \citep[e.g.][]{maeder83} is confirmed 
impressively by observation: when massive stars enter the Wolf-Rayet phase as 
WN subtypes, equilibrium CNO-processed material becomes exposed on their surface
\citep[see e.g.][]{crowther07}.

However, traces of mixing of CNO-cycled products from the stellar core
to the stellar surface can already be found much earlier in the lives of massive stars. 
Observational indications of superficial abundance anomalies for
carbon, nitrogen, and oxygen (and the burning product helium) in OB-type stars even on the main
sequence (MS) and, more prominently, in the blue supergiants
was found early in classification spectrograms \citep[e.g.][]{walborn76}. 
Subsequent analyses provided evidence for a characteristic
enrichment of nitrogen -- which is the easiest to be detected -- and helium in
many early-type stars, both near the MS and in blue supergiants  
\citep[e.g.][]{schoenberner88,gila92,herrero92,kilian92,venn95,lyubimkov96,mcerlean99}.

A theoretical understanding of {\em early} mixing of CNO-pro\-cessed
material to the stellar surface could not be achieved within the
framework of evolution calculations for non-rotating stars with
mass-loss, which were state-of-the-art at that time
\citep[e.g.][]{chma86}. The pollution of the surface layers with CNO-cycled 
material only occurs when the star reaches the red supergiant phase in
such models, via convective dredge--up. 

Considered a secondary effect for a long time, stellar rotation has come
lately into focus. It turned out that rotationally-induced mixing 
through meridional circulation and turbulent diffusion in rotating stars 
provides the means to change all model outputs substantially and to bring 
theory and observation into much better agreement \citep{mame00,hela00}. 
The latest step taken in the modelling was to consider the effects 
from an interplay of rotation and magnetic fields, which -- depending
on the detailed input physics and approximations made -- only provide minor 
modifications to the
surface abundances \citep{heger05} or substantial changes
\citep[henceforth abbreviated as MM05]{mame05}.
These differences result mainly from the two groups
using different sets of equations for computing the effects of magnetic  
fields; see in particular, the changes brought to the system of equations  
given by \citet{spruit02} by MM05 in their Sect.~2.

The only means to verify the models is via systematic comparison with
observations covering the relevant parts of the Hertzsprung-Russell diagram. 
Homogeneous analyses of a larger star sample from the main sequence to the
supergiant stage have only recently become available. 
The results in particular for N abundances 
apparently challenge the concept of rotational mixing in massive stars
\citep{hunter09} and thus the present-day evolution models
\citep[see, however,][]{maederetal09}.

\begin{figure*}
\centering
\hfill
\includegraphics[width=.485\textwidth]{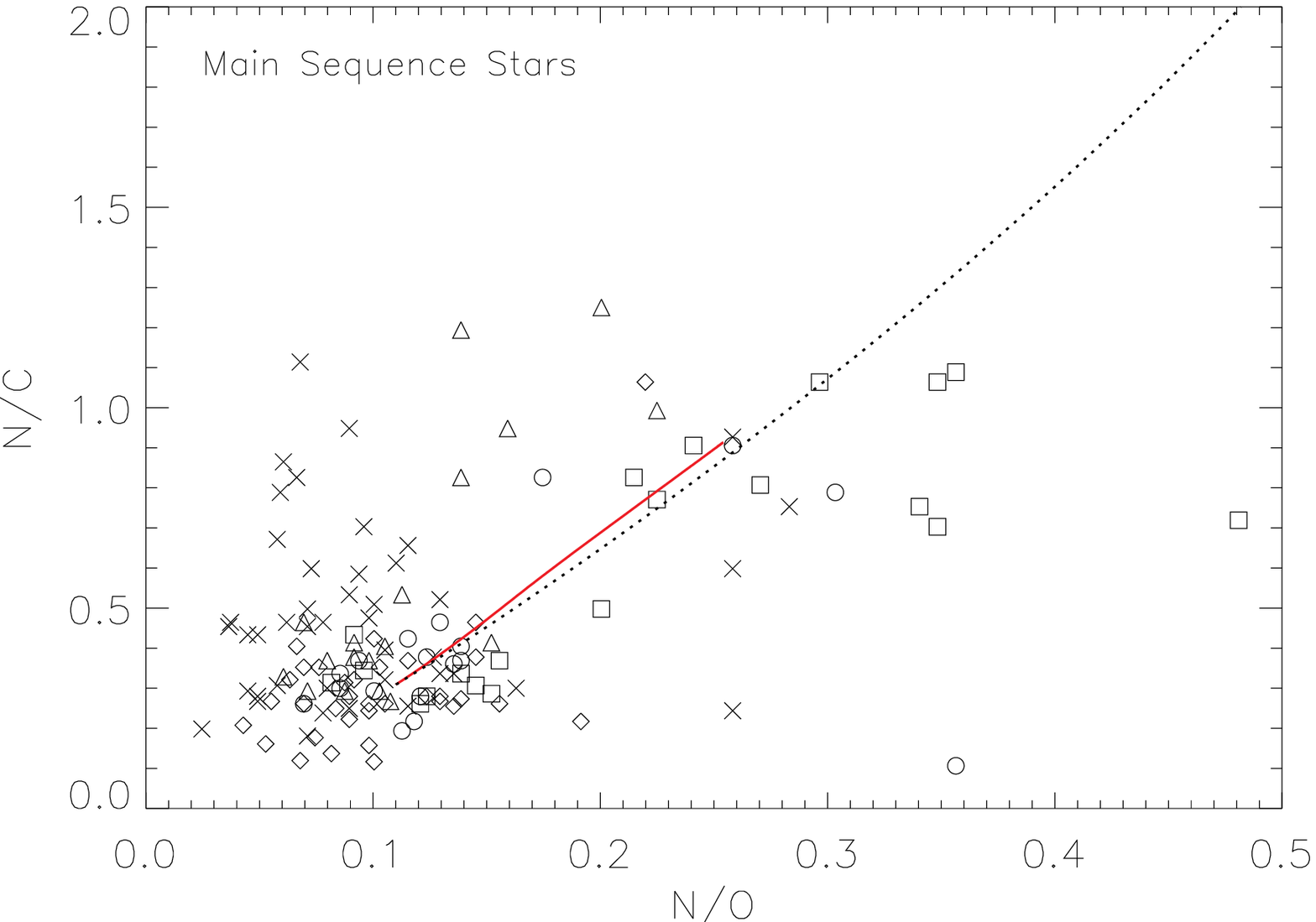}\hfill
\includegraphics[width=.485\textwidth]{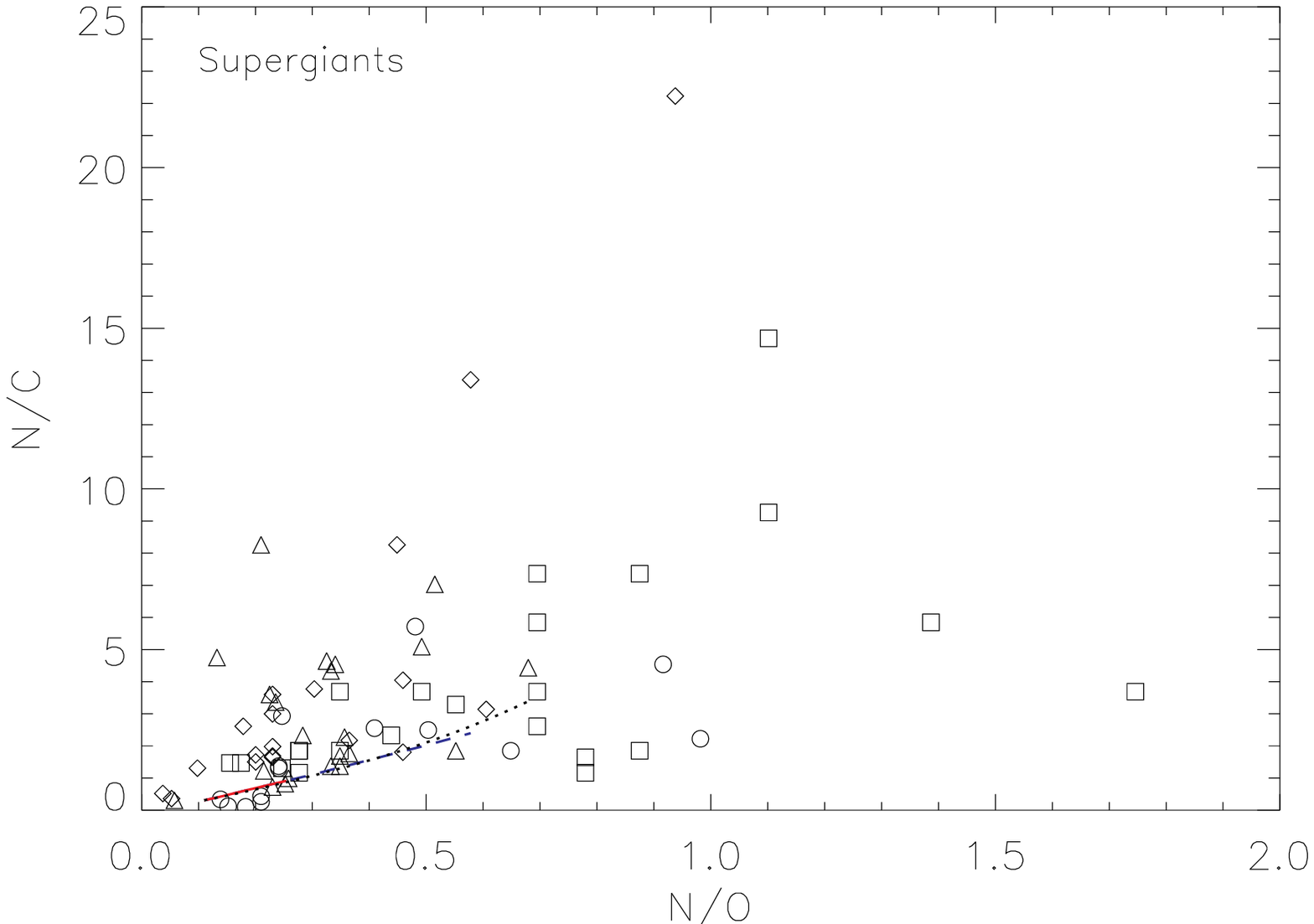}
\hfill
\caption{Status of observational constraints on the (magneto-)hydrodynamic mixing of 
CNO-burning products in massive stars from previous NLTE analyses. Mass ratios $N/C$
over $N/O$ are displayed. Left panel: main-sequence stars.
Circles: \citet{kilian92}; triangles: \citet{gila92}; 
diamonds: \citet{cula94}, \citet{daflon99,daflon01a,daflon01b}; 
squares: \citet{morel08}; crosses: \citet{hunter09}.
Right panel: BA supergiants.
Triangles: \citet{venn95}, \citet{vepr03};
circles: \citet{takeda00}; squares: \citet{crowther06}; diamonds: \citet{searle08}.
Error bars are omitted for clarity: uncertainties in the
abundances of the individual elements are typically about a factor 2, such that the
error bars can be larger than the plotting range.
The lines represent predictions from evolution
calculations, for a rotating 15\,$M_\odot$ star
\citep[$v_\mathrm{rot}^\mathrm{ini}$\,=\,300\,km\,s$^{-1}$,][MM03;
until the end of the MS: solid red line, until the end of He burning:
dashed blue line]{mema03} and for a star of the same mass 
and $v_\mathrm{rot}^\mathrm{ini}$ that in addition
takes the interaction of rotation and a magnetic dynamo into account
\citep[MM05; until the end of the MS: dotted
line]{mame05}, respectively.
The predicted trends are similar for the entire mass range
under investigation, see Fig.\ref{cnomix920}.} 
\label{litsummary}
\end{figure*}

Here, we address the topic of early mixing of CNO-cycled products in
massive stars from a fundamental perspective. We compare
the model predictions for the relative changes within the C:N:O
abundance ratios 
and the buildup 
of helium (Sect.~\ref{theory}) with observations.
For this we briefly review the status of the literature on CNO abundances
and introduce a well-studied Galactic star sample for which high-accuracy analyses 
of high-quality 
spectra were performed using improved NLTE modelling and comprehensive analysis
techniques (Sect.~\ref{obs}). Conclusions from this comparison are
drawn in Sect.~\ref{conclusions}.

\section{Theoretical considerations\label{theory}}
Diagnostic plots of an abundance ratio vs. another one, such as $N/C$ vs. $N/O$ or 
helium vs. $N/O$ discussed in this paper,
depend on both the changes produced by nuclear reactions and 
the dilution effects produced by mixing.  Let us estimate the slope
$\frac{{\rm d}(N/C)}{{\rm d}(N/O)}$
produced by the nuclear effects at the beginning of CNO burning, 
with everything expressed in mass fractions. Apart from the very massive stars 
($M$\,$>$\,40\,$M_{\odot}$), one may assume that, at the beginning of the burning, 
the $^{14}$N enhancement comes from the $^{12}$C destruction via the
CN cycle, and the oxygen content $O$ remains about constant. Thus, one has 
${\rm d}C= -\frac{6}{7} {\rm d}N $ since 
$^{14}$N globally results from the addition of two protons to $^{12}$C, 
\begin{eqnarray}
{\rm d}(N/O)={\rm d}N/O  \;, \nonumber \\ \vspace*{2mm}
{\rm d}(N/C)= \frac{{\rm d}N}{C} -\frac{N}{C^2} \, \frac{{\rm
d}C}{{\rm d}N} \,{\rm d}N=
\frac{{\rm d}N}{C}\left[1+ \frac{6}{7}\,\frac{N}{C} \right] \;.
\end{eqnarray}
\noindent
This gives  the slope
\begin{eqnarray}
\frac{{\rm d}(N/C)}{{\rm d}(N/O)}=
\frac{(N/C)}{(N/O)} \left[1+ \frac{6}{7}\, \frac{N}{C}\right] \;,
\label{slope}
\end{eqnarray}
\noindent
always in mass fractions. 
With  initial\footnote{
The chemical composition of the evolution
models is identical to those used in the OPAL opacity tables: the
solar mixture
of \cite{grno93}.}
ratios $N/C$ and $N/O$ of 0.31 and 0.11, respectively, we get
\begin{eqnarray}
\frac{{\rm d}(N/C)}{{\rm d}(N/O)}=3.77 \;.
\label{sl4} 
\end{eqnarray}
\noindent
This ratio is evidently greater than 1, since as $N$ starts growing, $C$ decreases, 
while $O$ does not vary much. The relation turns slightly upward as
$N/C$ is increasing owing to the term in brackets in Eq.~(\ref{slope}). However, 
at some advanced stage in evolution, corresponding to WN stars not shown here, 
the curve will saturate and turn down slightly \citep[p.~699]{maeder09}, since the 
CN cycle is then at equilibrium, while $^{16}$O is still turned to $^{14}$N.
Dilution mixes a fraction $f$ of $N$\,+\,$\Delta N$ enriched and C depleted materials 
with a fraction $(1-f)$ of the original $N$ and $C$. Under the same assumptions 
as above, it is easy to show that, to the first order, the slope for the 
relative enrichments in the $N/C$ vs. $N/O$ plot behaves the same way as in 
Eq.~(\ref{slope}) independently of $f$. 

The value of $f$, however, determines the amplitudes of the departures from the cosmic ratios. 
Our models with rotational mixing (Meynet \& Maeder~2003, MM03; 
Ekstr\"om et al.~2008, E08)
or with rotation and magnetic fields 
(MM05), as illustrated e.g. in Fig.~\ref{litsummary},
have an initial slope $\frac{{\rm d}(N/C)}{{\rm
d}(N/O)}$\,$\approx$\,4,
which is in excellent agreement with Eq.~(\ref{sl4}). 
The amplitude $f$ of the mixing depends on the various model assumptions, in particular 
on the treatment of the shear mixing with or without horizontal turbulence. 
The models without horizontal turbulence \citep[MM00]{mema00}
predict more mixing than models that account for it (MM03).
Models that include both rotation and magnetic field predict a still larger  
mixing (MM05). 

Let us now consider the behaviour of the helium surface content 
$Y_{\mathrm{s}}$ vs. $N/O$ (as illustrated later).
Strictly and only at the very beginning 
of the CN burning, and under the assumption of an initially constant 
oxygen, we get ${\rm d}Y_{\mathrm{s}}$\,$=$\,$\frac{2}{7}{\rm d}N$, since 
when 4 units of mass of helium are made, 14 units of mass of nitrogen are produced.
The slope is 
\begin{eqnarray}
\frac{{\rm d}Y_{\mathrm{s}}}{{\rm d}(N/O)} = \frac{2}{7}\, O 
\approx 0.286 \times 0.009=0.0026\; ,
\end{eqnarray}
i.e., it is essentially flat initially. Later in the evolution, both N and O 
change simultaneously, and one has to rely on numerical 
models. The resulting slope in the models can vary; e.g., a steeper slope is 
inferred for the 25\,$M_{\odot}$ model of both MM03 and models by MM00
than for the 15\,$M_{\odot}$ models of MM03 and MM05. This depends on
whether the matter that arrives at the surface comes from inner
regions that are at both CN and ON equilibria, or only at CN equilibrium.
There are, of course, a range of intermediate cases.

\begin{table*}
\caption{Stellar parameters and elemental abundances of the sample stars}
\label{table1}
\centering
\begin{tabular}{lrrrrrrrrr}
\hline\hline
Star & $T_\mathrm{eff}$ & $\log g$ & $v \sin i$    & 
$Y_\mathrm{S}$ & $\varepsilon$\,(C)$^1$ & $\varepsilon$\,(N) & $\varepsilon$\,(O) & $\varepsilon$\,($\sum\mathrm{CNO}$) & $M/M_\odot$ \\
     &              (K) &    (cgs) & (km\,s$^{-1}$)& (by mass)\\
\hline
\multicolumn{3}{l}{B-type main-sequence stars:}\\
\object{HD36591}  & 27000$\pm$300 & 4.12$\pm$0.05 & 12$\pm$1 & 0.28$\pm$0.03 & 8.33$\pm$0.08 & 7.75$\pm$0.09 & 8.75$\pm$0.11 & 8.92$\pm$0.07 & 13\\
\object{HD61068}  & 26300$\pm$300 & 4.15$\pm$0.05 & 14$\pm$2 & 0.28$\pm$0.03 & 8.27$\pm$0.07 & 8.00$\pm$0.12 & 8.75$\pm$0.09 & 8.93$\pm$0.06 & 12\\
\object{HD63922}  & 31200$\pm$300 & 3.95$\pm$0.05 & 29$\pm$4 & 0.25$\pm$0.03 & 8.34$\pm$0.08 & 7.77$\pm$0.08 & 8.79$\pm$0.10 & 8.95$\pm$0.07 & 19\\
\object{HD74575}  & 22900$\pm$300 & 3.60$\pm$0.05 & 11$\pm$2 & 0.28$\pm$0.03 & 8.37$\pm$0.10 & 7.92$\pm$0.10 & 8.80$\pm$0.08 & 8.98$\pm$0.06 & 12\\
\object{HD122980} & 20800$\pm$300 & 4.22$\pm$0.05 & 18$\pm$1 & 0.28$\pm$0.03 & 8.32$\pm$0.09 & 7.76$\pm$0.08 & 8.72$\pm$0.05 & 8.90$\pm$0.04 &  8\\
\object{HD149438} & 32000$\pm$300 & 4.30$\pm$0.05 &  4$\pm$1 & 0.28$\pm$0.03 & 8.30$\pm$0.12 & 8.16$\pm$0.12 & 8.77$\pm$0.08 & 8.97$\pm$0.06 & 17\\[1mm]
\multicolumn{3}{l}{BA-type supergiants:}\\                                                                                            
\object{HD13476}  &  8500$\pm$150 & 1.40$\pm$0.10 & 12$\pm$2 & 0.36$\pm$0.08 & 8.18$\pm$0.11 & 8.58$\pm$0.04 & 8.63$\pm$0.06 & 8.98$\pm$0.03 & 16\\
\object{HD14433}  &  9150$\pm$150 & 1.40$\pm$0.10 & 17$\pm$3 & 0.32$\pm$0.07 & 8.23$\pm$0.04 & 8.23$\pm$0.03 & 8.67$\pm$0.05 & 8.91$\pm$0.03 & 19\\
\object{HD34085}  & 12100$\pm$150 & 1.75$\pm$0.10 & 25$\pm$3 & 0.32$\pm$0.04 & 8.23$\pm$0.09 & 8.47$\pm$0.06 & 8.75$\pm$0.05 & 9.01$\pm$0.03 & 23\\
\object{HD87737}  &  9600$\pm$150 & 2.00$\pm$0.10 &  2$\pm$2 & 0.37$\pm$0.03 & 8.25$\pm$0.06 & 8.52$\pm$0.08 & 8.73$\pm$0.06 & 9.02$\pm$0.04 & 10\\
\object{HD92207}  &  9500$\pm$200 & 1.20$\pm$0.10 & 30$\pm$5 & 0.35$\pm$0.06 & 8.33$\pm$0.08 & 8.25$\pm$0.04 & 8.79$\pm$0.07 & 9.00$\pm$0.04 & 27\\
\object{HD111613} &  9150$\pm$150 & 1.45$\pm$0.10 & 17$\pm$2 & 0.35$\pm$0.05 & 8.29$\pm$0.10 & 8.46$\pm$0.04 & 8.73$\pm$0.04 & 9.01$\pm$0.03 & 17\\
\object{HD187983} &  9300$\pm$250 & 1.60$\pm$0.15 & 15$\pm$6 & 0.32$\pm$0.05 & 8.29$\pm$0.09 & 8.19$\pm$0.04 & 8.78$\pm$0.02 & 8.98$\pm$0.02 & 15\\
\object{HD195324} &  9200$\pm$150 & 1.85$\pm$0.10 &  3$\pm$3 & 0.38$\pm$0.06 & 8.10$\pm$0.11 & 8.70$\pm$0.08 & 8.73$\pm$0.04 & 9.07$\pm$0.04 & 11\\ 
\object{HD197345} &  8700$\pm$150 & 1.20$\pm$0.10 & 10$\pm$9 & 0.37$\pm$0.04 & 8.09$\pm$0.07 & 8.56$\pm$0.07 & 8.69$\pm$0.04 & 8.99$\pm$0.03 & 22\\
\object{HD202850} & 10800$\pm$200 & 1.85$\pm$0.10 & 14$\pm$5 & 0.38$\pm$0.06 & 8.16$\pm$0.04 & 8.70$\pm$0.06 & 8.76$\pm$0.05 & 9.09$\pm$0.03 & 15\\
\object{HD207673} &  9250$\pm$100 & 1.80$\pm$0.10 &  1$\pm$2 & 0.33$\pm$0.08 & 8.17$\pm$0.09 & 8.48$\pm$0.03 & 8.72$\pm$0.04 & 8.99$\pm$0.03 & 12\\ 
\object{HD210221} &  8400$\pm$150 & 1.40$\pm$0.10 &  0$\pm$0 & 0.34$\pm$0.03 & 8.22$\pm$0.06 & 8.52$\pm$0.06 & 8.70$\pm$0.05 & 9.00$\pm$0.03 & 15\\
\object{HD212593} & 11200$\pm$200 & 2.10$\pm$0.10 &  6$\pm$2 & 0.35$\pm$0.05 & 8.30$\pm$0.08 & 8.44$\pm$0.06 & 8.74$\pm$0.04 & 9.01$\pm$0.03 & 12\\
\object{HD213470} &  8400$\pm$150 & 1.30$\pm$0.10 & 13$\pm$1 & 0.32$\pm$0.07 & 8.17$\pm$0.08 & 8.54$\pm$0.04 & 8.65$\pm$0.03 & 8.97$\pm$0.02 & 18\\
\hline\\[-5mm]
\end{tabular}
\begin{list}{}{}
\item [$^1$] $\varepsilon\,(X)$\,$=$\,$\log$\,$X$/H\,+\,12
\end{list}
\end{table*}

\section{Observational constraints\label{obs}}
Numerous studies of CNO abundances in massive stars of the Milky Way are available from
the literature, mostly for early B-type stars close to the MS and for B and A-type
supergiants. We illustrate the results of several key publications in the $N/O$--$N/C$ diagrams of Fig.~\ref{litsummary}. Some of the more
recent studies \citep{crowther06,searle08,hunter09} are based on NLTE
model atmospheres, while the bulk of the data were obtained from NLTE
line-formation computations on LTE model atmospheres -- which is equivalent 
to the full NLTE approach in the cases under consideration \citep[NP07]{nipr07}.

The MS stars show a concentration around the solar ratios of $N/C$\,$\approx$\,0.3 and
$N/O$\,$\approx$\,0.1. However, overall a wide range of $N/O$--$N/C$
combinations has been realised, with the deviations from the predictions
increasing in the supergiants. On the one hand, this leaves room for broad
interpretation -- even more so if only one of the elements (like
N) is considered -- including the statement that some
observational data points pose 
a challenge for the evolution models. On the other hand, most of
the data are nevertheless consistent with the predictions, as the abundance
uncertainties are very large. Typically, the statistical 1$\sigma$-error in
abundance per element is about a factor $\sim$2, and systematic
uncertainties are often largely underestimated \citep[for a
discussion of this see][]{nipr10} 
or even unaccounted for. The error bars in Fig.~\ref{litsummary} are
larger than the entire plotting range in many cases. In consequence,
no {\em definite} conclusions can be drawn on the quality of the stellar
evolution models from these data.

In past years a number of high-accuracy studies of massive stars
in the solar neighbourhood 
have been published by us. Carefully analysed data are available on 6 slowly-rotating early
B-type stars near the MS \citep[PNB08]{nipr06,nipr07,nipr08,przybilla08}
and on 14 BA-type supergiants \citep{przybilla06,firnstein06,schpr08}. 

\begin{figure}
\centering
\includegraphics[width=.485\textwidth]{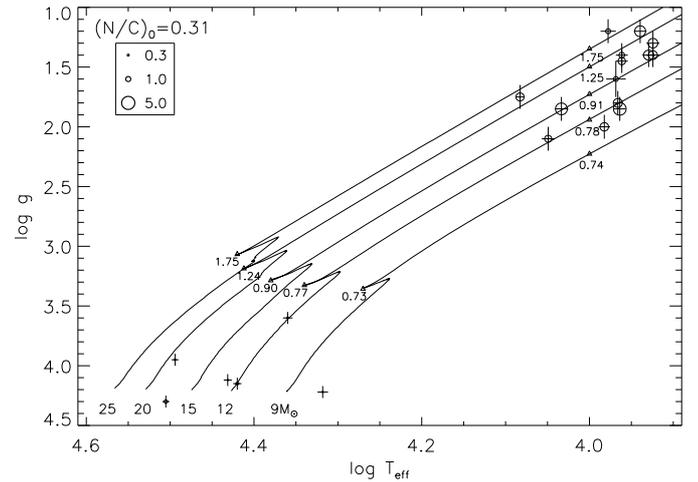}
\caption{Sample stars in the $\log T_\mathrm{eff}$--$\log
  g$-plane. Diamonds denote MS stars and circles supergiants.
  The symbol size encodes the $N/C$ ratio (by mass), with
  typical values exemplified in the legend. Evolutionary tracks
  for rotating stars of metallicity $Z$\,=\,0.02 (MM03) are shown. 
  Predicted $N/C$ ratios are indicated along
  the tracks.} 
\label{tefflogg}
\end{figure}

\begin{figure}
\centering
\includegraphics[width=.488\textwidth]{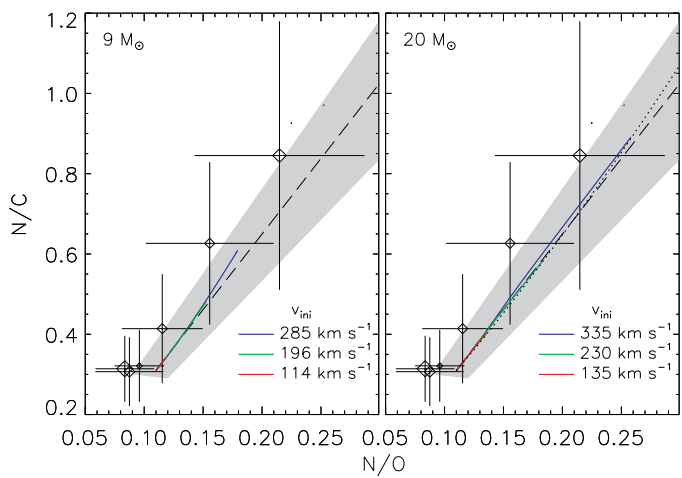}\\[-2mm]
\caption{Comparison of $N/C$ vs. $N/O$ abundance ratios (by mass) for our 
six MS sample stars (diamonds) with model predictions for 9\,$M_\odot$
(left) and 20\,$M_\odot$ stars (right panel) at different
$v_\mathrm{rot}^\mathrm{ini}$ (see legend, equivalent to 30, 50 and 70\%
of the star's breakup velocity, E08). The dotted line (right
panel only) describes the
magnetic 15\,$M_{\odot}$ model of MM05 with
$v_\mathrm{rot}^\mathrm{ini}$\,=\,300\,km\,s$^{-1}$.
The symbol size encodes the stellar
mass and error bars give 1$\sigma$-uncertainties. The long-dashed lines
correspond to the theoretical slope 3.77 as deduced in Sect.~\ref{theory}.
The grey area spans the full range of theoretical slopes using
different references for solar abundances \citep[see text, lower envelope
defined by data of][]{asplund09} and the cosmic abundance
standard of PNB08 (upper envelope).}
\label{cnomix920}
\end{figure}

\begin{figure}
\centering
\includegraphics[width=.488\textwidth]{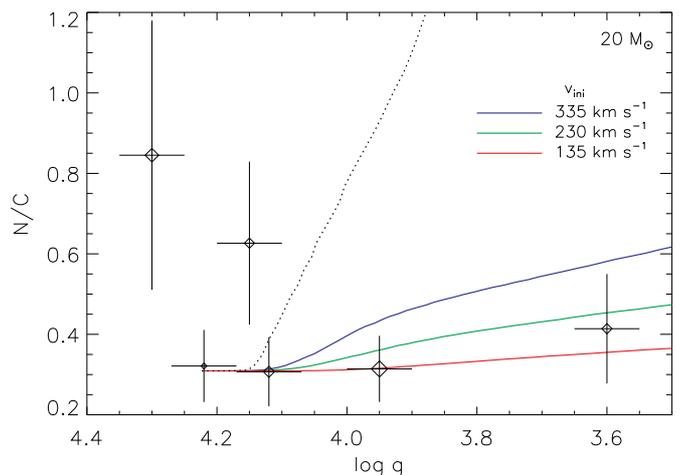}\\[-2mm]
\caption{$N/C$ abundance ratios on the MS as function
of $\log g$. The same tracks and observations as in
Fig.~\ref{cnomix920} (right panel) are displayed.}
\label{cnomix_gnc}
\end{figure}

\begin{figure}
\centering
\includegraphics[width=.488\textwidth]{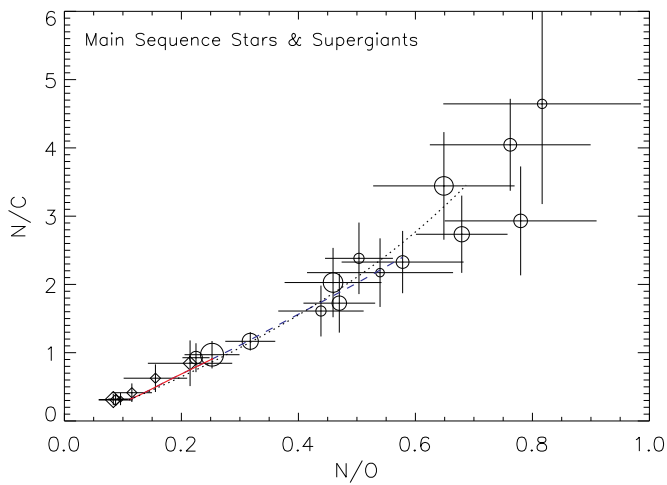}
\caption{$N/C$ vs. $N/O$ abundance ratios (by mass) for our sample stars.
  B-type MS stars are displayed as diamonds, BA-type supergiants 
  as circles. The symbol size encodes the stellar mass and error bars 
  give 1$\sigma$-uncertainties. The different lines describe 
  model predictions for 15\,$M_{\odot}$ stars identical to 
  those in Fig.~\ref{litsummary}.}
\label{ourcnomix}
\end{figure}

\begin{figure}
\centering
\includegraphics[width=.488\textwidth]{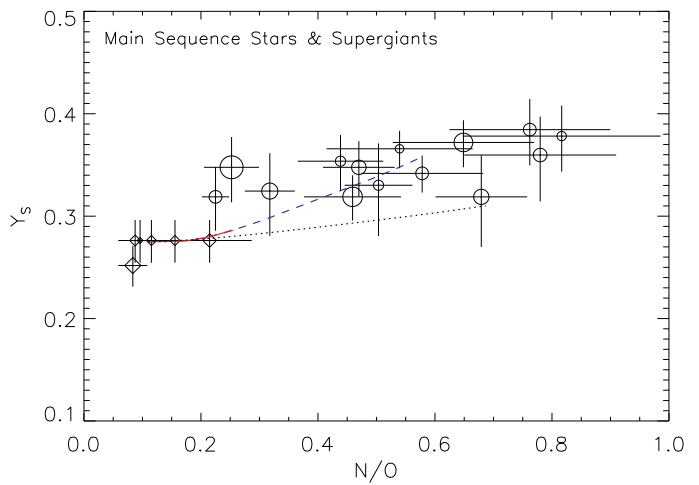}
\caption{Surface helium abundance $Y_\mathrm{S}$ as a function of $N/O$ ratio (both by mass) 
  for our sample stars. Symbol and line encoding as in
  Fig.~\ref{litsummary}.}
\label{hemix}
\end{figure}

In brief, high-resolution ($R$\,=\,40--48\,000) and high-S/N
spectra ($S/N$\,$>$\,300) with wide wavelength coverage and thorough
continuum normalisation (obtained with FOCES@Calar Alto 
Observatory and FEROS@ESO/La Silla) were analysed using a hybrid NLTE approach
(Przybilla et al. 2006; NP07). State-of-the-art atomic input data were
used in the modelling. In contrast to all previous work, multiple hydrogen 
lines, the helium lines, multiple metal ionisation
equilibria and the stellar energy distributions were reproduced
{\em simultaneously} in an iterative approach to determine the stellar
atmospheric parameters. 
Chemical abundances were derived from analysis of practically the entire
observable spectrum per element. The rewards of such a comprehensive,
but time-consuming procedure are unprecedentedly small statistical error margins and largely
reduced systematics. The relevant results -- effective temperature 
$T_\mathrm{eff}$, surface gravity $\log g$, 
projected rotational velocity $v \sin i$,
surface helium abundance
$Y_\mathrm{S}$, CNO abundances and their total, and the zero-age MS mass 
(derived under the assumption that the objects have evolved
directly from the MS)
-- are summarised in Table~\ref{table1} 
and visualised in
Figs.~\ref{tefflogg}--\ref{hemix}. We re-ran the
analysis of the supergiant targets for the present work, taking
advantage of improved model grids.
Our new results for these supergiants
agree with the earlier ones within the uncertainties but are more accurate.

The sample is displayed in the $\log T_\mathrm{eff}$--$\log g$-plane
and compared to evolutionary tracks for rotating 
stars\footnote{The metal content (by mass) of the models is $Z$\,$=$\,0.020,
while a value of 0.014 seems more appropriate for the massive star
population in the solar neighbourhood (PNB08). No fundamental changes to 
our conclusions are expected to arise from this difference. 
The tracks will shift a bit, in particular the zero-age MS towards higher gravities.}
in Fig.~\ref{tefflogg}. 
The stars have initial masses between about 9 to 25\,$M_\odot$.
Stars close to the MS are apparently slow rotators, while
the atmospheric expansion has erased any indication of the initial
rotational velocity in the supergiants. There is good qualitative
agreement between the observations and predictions for the $N/C$ ratios,
finding low values close to the initial (solar) ratio
near the MS and enhanced values in the supergiants. However, the
observed $N/C$ ratios reach much higher values than predicted by the models 
discussed in Fig.~\ref{tefflogg}, 
provided these stars have
recently left the MS and evolve now into red supergiants.

The behaviour of light element abundances of the MS sample stars is
shown in Fig.~\ref{cnomix920}. 
Theoretical evolutionary tracks are
overplotted. It is striking to see that, regardless of the initial mass and
of the initial rotation, the slope of the theoretical tracks
is always the same and equal to the nuclear effect of the CN cycle as
derived analytically in Sect.~\ref{theory}\footnote{The theoretical slope varies
between $\sim$3.0 and 4.2 for typical initial CNO abundances
\citep[e.g. solar values according to][or the cosmic abundance
standard of PNB08]{angr89,grsa98,asplund09}, indicated by the grey area in
Fig.~\ref{cnomix920}. Improved agreement between observation and
theory may be achieved when tailored abundances are used for the
stellar evolution models, see also the previous footnote.}. Thus the observed slope is 
a confirmation of the activity of the CN cycle. It does not depend on the 
stellar model but constitutes observational evidence of the nuclear path.
As already indicated before, what is model dependent are the
amplitudes of the departure from the cosmic ratios.
Figure~\ref{cnomix920} shows that the observed points may be reproduced
from models having differing velocities and masses, with and without
magnetic field;
however, by itself this diagram cannot help in disentangling the
various possibilities.

The same evolutionary tracks are plotted in the $N/C$ vs. $\log g$
diagram in Fig.~\ref{cnomix_gnc}. We see that, except for the stars HD\,61068 and
HD\,149438 \citep[$\tau$\,Sco, which is a genuine slow rotator with a
rather strong magnetic field\footnote{A magnetic field is present also
in HD\,74575 \citep{hubrig09}.} of probably fossil origin,][]{donati06}, the other points may 
be accounted for by models having low initial rotation. HD\,61068 is marginally
compatible with the 15\,$M_\odot$ model 
at $v_\mathrm{rot}^\mathrm{ini}$\,=\,300\,km\,s$^{-1}$
computed with magnetic field, while
$\tau$\,Sco challenges present stellar models. It has a behaviour
that may be explained by a homogeneous evolution, but this still has
to be confirmed by further computations. This is a very
interesting star that certainly deserves further inspections both from the
observational and theoretical points of view.

The behaviour of light element abundances in the whole star sample is shown in
Fig.~\ref{ourcnomix}. In contrast to the literature values
(Fig.~\ref{litsummary}), a clear and tight trend
is found, confirming the predicted locus of $N/O$--$N/C$ abundance ratios.
However, as already indicated above, the MM03 models for rotating stars with
mass loss evolving towards the red supergiant stage (solid line in
Fig.~\ref{ourcnomix}) 
predict mixing that is too low, i.e. too low $f$ (Sect.~\ref{theory}),
in particular for most of the supergiants. 
A combination of five reasons may provide an explanation.\\[1mm]
{\sc i}) Higher than average rotation velocities in the progenitor stars of these
supergiants on the MS may reconcile the situation for some objects.\\
{\sc ii}) Evolution models for rotating stars that also account
for the interaction of rotation and a magnetic dynamo (MM05)
predict enhanced mixing signatures of the amount required (dotted
line).\\ 
{\sc iii}) Some stars may have evolved in a close binary, which can also 
lead to enhanced mixing associated with mass transfer.\\
{\sc iv}) Some objects may have been siblings to $\tau$\,Sco on the
MS, climbing up the $N/O$--$N/C$ relation even further in their further evolution.\\
{\sc v}) Supergiants may already have evolved through the red
supergiant phase (e.g., on a blue loop) to expose first dredge-up abundance
ratios, which could quantitatively also explain the observations (dashed
line). 

More information may be derived from the helium content,
which in the case of BA-type supergiants is determined here for a significant 
number of stars for the first time in a self-consistent analysis. Our
results are displayed in Fig.~\ref{hemix}. 
On the MS no helium surface enrichment is observed, as predicted
in the models for stars with masses below about 20\,$M_\odot$.
After the MS, the picture is blurred by the possible occurrence of a blue
loop.
Actually the interpretation of the blue supergiant can become
really constraining only when we obtain additional hints
to the previous evolution of the star. Has the blue supergiant
evolved directly from the MS, or has it evolved in that stage
after going through a red supergiant stage?
At the moment, from the models the situation would be the following:
for models below 20\,$M_\odot$,
He-enrichments at the level observed in the present supergiants are
only compatible with models having undergone a dredge-up in the red
supergiant phase. This is true whether rotation is considered or not,
or a magnetic field is accounted for or not. The present
track for the magnetic 15\,$M_\odot$ model was computed only up to the
end of the MS phase and thus did not yet go through the dredge-up phase.
Depending on the rotation velocity, the presence of a magnetic field
or its absence, models, after the red supergiant phase, will populate 
diverse parts of the region in the plane $Y_{\rm S}$ versus $N/O$,
as illustrated e.g. by the dashed and dotted lines in Fig.~\ref{hemix}.

We note, however, that we cannot exclude at present the possibility that the observed helium
abundances in the supergiants may be overestimated. A
systematic downward shift by a mere 10-15\% (which is within the
typical systematic uncertainties in our abundance determinations) 
would be sufficient, e.g., to bring the observations and the magnetic
model in Fig.~\ref{hemix} into agreement.
All observed lines arise from two energetically close \ion{He}{i} levels only 
(2p\,\element[\circ][1]{P},\element[\circ][3]{P}). Model
atom shortcomings (such as insufficient {\em
ab-initio} collisional data),
which may become important only in the weak-line limit at the cool
temperature border, could therefore remain unnoticed and could give
rise to systematics. It would not be the first
time that uncertainties in atomic data complicate the statistic equilibrium
and radiative transfer calculations needed to interpret the observed
\ion{He}{i} lines \citep[e.g.][]{przybilla05,najarro06}. Further
investigations are required before firm conclusions on the
evolutionary state of the supergiants are drawn from the helium
abundances, and a fully coherent picture can be deduced.

Finally, an inspection of the total of the CNO abundances 
(see Table~\ref{table1}) is instructive. Because the CNO cycles are 
cataly\-tic, the sum of particles is conserved. Indeed, the sample shows only 
a small scatter around the mean
combined abundance, $\log \sum$CNO\,$+$\,12\,=\,8.98$\pm$0.05. This
is further (indirect) evidence that chemical inhomogeneities among the massive
star population in the solar neighbourhood are small
(Nieva \& Przybilla~2008; PNB08), even at extended distances out to 1--2\,kpc from the Sun.
Predictions of a high efficiency of hydrodynamic mixing in the
interstellar medium and consequently of a locally well-mixed 
Milky Way \citep{edmunds75,roku95} are thus supported as well.

\section{Conclusions\label{conclusions}}
Evolution models for massive stars predict very tight relations for
the change of surface $N/O$ and $N/C$ abundance ratios and for the
buildup of helium, as a consequence of mixing with CNO-cycled matter
(for given initial chemical composition).  
Massive stars in the solar neighbourhood, which supposedly share
a nearly uniform initial composition (PNB08), are
therefore expected to follow the predicted relations, which are
governed by nuclear reactions and the dilution effects
produced by mixing.

A comparison with NLTE spectral analyses from the literature
on CNO abundances in massive stars of $\sim$8--25\,$M_\odot$
leaves room for broad interpretation because of large uncertainties.
Observations may even be interpreted as posing a challenge to theory,
in particular when the full set of information is not accounted for
and conclusions are drawn from one indicator alone, e.g. nitrogen.

On the other hand, our high-precision analyses of a sample of Galactic massive
stars from the main sequence to the supergiant stage  
find these tight correlations. Even though further investigations are required to
refine the observational constraints on helium abundances before a
fully coherent picture can be obtained, the case for strong mixing is clearly supported. 
It is predicted, e.g., by the models of MM00, by models with magnetic field
(MM05), or in cases where the models have gone through the first dredge-up. 
The separation of the different possibilities may come from studing the 
evolution of the rotation velocities with time. On the one hand,
models with magnetic field predict higher
rotational velocities at the end of the MS phase, because the strong 
internal coupling transmits some of the fast core rotation to the
surface. On the other hand, stars on a blue loop should be slower rotators than
objects on their first passage from blue to red on average
because of the additional angular-momentum loss experienced through
strong mass loss during the red supergiant phase. Indeed, some of the
slowest rotators show very strong mixing signatures (Table~\ref{table1}).
 
Finally, we have to point out that some of our results have the
potential of challenging the currently available evolution models.
The star $\tau$\,Sco (HD149438) stands out in the sample as it shows
characteristics that may be explained by a homogeneous evolution, but
it requires a highly-efficient spin-down mechanism. One may speculate on
magnetic breaking due to angular-momentum losses by a
magnetically confined line-driven stellar wind or magnetic coupling
to the accretion disc during the star-forming process in the case of a fossil field. 
Even though the topic is not understood theoretically in a comprehensive
way, spin-down times of the order of 1\,Myr \citep{ud-Doula09} or even less
\citep{mikulasek08} are reported for some magnetic massive stars, 
possibly leading to the required slow rotation already on the 
zero-age MS in this case as well. Then, if the present
helium abundances are confirmed and the supergiants are shown to have
evolved directly from the MS, a different kind of mixing may also be
required.

We conclude that the tight observational constraints that are required for a 
thorough testing of the stellar evolution models 
are within reach.
Applications of the improved modelling and analysis techniques to
high-quality observations of larger star samples in the Milky Way and other galaxies 
may finally provide the empirical basis to benchmark the models.
It may thus become feasible to disentangle the effects of
metallicity, rotation, magnetic fields, and binarity 
on massive star evolution.

\begin{acknowledgements}
NP and MFN would like to thank the staff at Geneva Observatory for
their hospitality. MF acknowledges funding by the
\emph{Deut\-sche For\-schungs\-ge\-mein\-schaft, DFG\/} under project
number PR\,685/3-1. Travel to the Calar Alto Observatory/Spain was
supported by \emph{DFG} under grant PR\,685/1-1. 

\end{acknowledgements}


\begin{thebibliography}{}

\bibitem[Anders \& Grevesse(1989)]{angr89}
 Anders, E., \& Grevesse, N. 1989, \gca, 53, 197

\bibitem[Asplund et al.(2009)]{asplund09}
 Asplund, M., Grevesse, N., Sauval, A. J., \& Scott, P. 2009, \araa, 47, 481

\bibitem[Chiosi \& Maeder(1986)]{chma86}
 Chiosi, C., \& Maeder, A. 1986, \araa, 24, 329

\bibitem[Crowther(2007)]{crowther07}
 Crowther, P. A. 2007, \araa, 45, 177

\bibitem[Crowther et al.(2006)]{crowther06}
 Crowther, P. A., Lennon, D. J., \& Walborn, N. R. 2006, \aap, 446, 279

\bibitem[Cunha \& Lambert(1994)]{cula94}
 Cunha, K., \& Lambert, D. L. 1994, \apj, 426, 170

\bibitem[Daflon et al.(1999)]{daflon99}
 Daflon, S., Cunha, K., \& Becker, S. R. 1999, \apj, 522, 950 

\bibitem[Daflon et al.(2001a)]{daflon01a}
 Daflon, S., Cunha, K., Becker, S. R., \& Smith, V. V. 2001a, \apj, 552, 309

\bibitem[Daflon et al.(2001b)]{daflon01b}
 Daflon, S., Cunha, K., Butler, K., \& Smith, V. V. 2001b, \apj, 563, 325

\bibitem[Donati et al.(2006)]{donati06}
 Donati, J.-F., Howarth, I. D., Jardine, M. M., et al. 2006, \mnras, 370, 629

\bibitem[Edmunds(1975)]{edmunds75}
 Edmunds, M. G. 1975, \apss, 32, 483

\bibitem[Ekstr\"om et al.(2008)]{ekstroem08}
 Ekstr\"om, S., Meynet, G., Maeder, A., \& Barblan, F. 2008, \aap,
 478, 467 (E08)

\bibitem[Firnstein(2006)]{firnstein06}
 Firnstein, M. 2006, Diploma Thesis, Univ. Erlangen-Nuremberg

\bibitem[Gies \& Lambert(1992)]{gila92}
 Gies, D. R., \& Lambert, D. L. 1992, \apj, 387, 673

\bibitem[Grevesse \& Noels(1993)]{grno93}
 Grevesse, N., \& Noels, A. 1993, in Origin and Evolution of the
  Elements, ed. S.~Kubono, \& T.~Kajino, 15

\bibitem[Grevesse \& Sauval(1998)]{grsa98}
 Grevesse, N., \& Sauval, A. J. 1998, \ssr, 85, 161

\bibitem[Heger \& Langer(2000)]{hela00}
 Heger, A., \& Langer, N.~2000, \apj, 544, 1016

\bibitem[Heger et al.(2005)]{heger05}
 Heger, A., Woosley, S.E., \& Spruit, H.C.~2005, \apj, 626, 350

\bibitem[Herrero et al.(1992)]{herrero92}
 Herrero, A., Kudritzki, R. P., Vilchez, J. M., et al. 1992, \aap, 261, 209

\bibitem[Hubrig et al.(2009)]{hubrig09}
 Hubrig, S., Briquet, M., De Cat, P., et al. 2009, Astron. Nachr., 330, 317

\bibitem[Hunter et al.(2009)]{hunter09}
 Hunter, I., Brott, I., Langer, N., et al. 2009, \aap, 496, 841

\bibitem[Kilian(1992)]{kilian92}
 Kilian, J. 1992, \aap, 262, 171

\bibitem[Lyubimkov(1996)]{lyubimkov96}
 Lyubimkov, L. S. 1996, \apss, 243, 329

\bibitem[Maeder(1983)]{maeder83}
 Maeder, A. 1983, \aap, 120, 113

\bibitem[Maeder(2009)]{maeder09}
 Maeder, A. 2009, Physics, Formation and Evolution of Rotating Stars (Springer Verlag, Berlin)

\bibitem[Maeder \& Meynet(2000)]{mame00}
 Maeder, A., \& Meynet, G.~2000, \araa, 38, 143

\bibitem[Maeder \& Meynet(2005)]{mame05}
 Maeder, A., \& Meynet, G.~2005, \aap, 440, 1041 (MM05)

\bibitem[Maeder et al.(2009)]{maederetal09}
 Maeder, A., Meynet, G., Ekstr\"om, S., \& Georgy, C. 2009, CoAst, 158, 72

\bibitem[McErlean et al.(1999)]{mcerlean99}
 McErlean, N. D., Lennon, D. J., \& Dufton, P. L. 1999, \aap, 349, 553

\bibitem[Meynet \& Maeder(2000)]{mema00}
 Meynet, G., \& Maeder, A. 2000, \aap, 361, 101 (MM00)

\bibitem[Meynet \& Maeder(2003)]{mema03}
 Meynet, G., \& Maeder, A. 2003, \aap, 404, 975 (MM03)

\bibitem[Mikul\'a\v{s}ek et al.(2008)]{mikulasek08}
 Mikul\'a\v{s}ek, Z., Krti\v{c}ka, J., Henry, G. W., et al. 2008, \aap, 485, 585

\bibitem[Morel et al.(2008)]{morel08}
 Morel, T., Hubrig, S., \& Briquet, M. 2008, \aap, 481, 453

\bibitem[Najarro et al.(2006)]{najarro06}
 Najarro, F., Hillier, D. J., Puls, J., Lanz, T., \& Martins, F. 2006, \aap, 456, 659

\bibitem[Nieva \& Przybilla(2006)]{nipr06}
 Nieva, M. F., \& Przybilla, N. 2006, \apj, 639, L39

\bibitem[Nieva \& Przybilla(2007)]{nipr07}
 Nieva, M. F., \& Przybilla, N. 2007, \aap, 467, 295 (NP07)
   
\bibitem[Nieva \& Przybilla(2008)]{nipr08}
 Nieva, M. F., \& Przybilla, N. 2008, \aap, 481, 199

\bibitem[Nieva \& Przybilla(2010)]{nipr10}
 Nieva, M. F., \& Przybilla, N. 2010, in ASP Conf. Ser., 
 in press (arXiv:0902.2949)

\bibitem[Przybilla(2005)]{przybilla05}
 Przybilla, N. 2005, \aap, 443, 293

\bibitem[Przybilla et al.(2006)]{przybilla06}
 Przybilla, N., Butler, K., Becker, S. R., \& Kudritzki R. P. 2006, \aap, 445, 1099

\bibitem[Przybilla et al.(2008)]{przybilla08}
 Przybilla, N., Nieva, M. F., \& Butler K. 2008, \apj, 688, L103 (PNB08)
 
\bibitem[Roy \& Kunth(1995)]{roku95}
 Roy, J.-R., \& Kunth, D. 1995, \aap, 294, 432
 
\bibitem[Schiller \& Przybilla(2008)]{schpr08}
 Schiller, F., \& Przybilla, N. 2008, \aap, 479, 849 

\bibitem[Sch\"onberner et al.(1988)]{schoenberner88}
 Sch\"onberner, D., Herrero, A., Becker, S. R., et al. 1988, \aap, 197, 209

\bibitem[Searle et al.(2008)]{searle08}
 Searle, S. C., Prinja, R. K., Massa, D., \& Ryans, R. 2008, \aap, 481, 777

\bibitem[Spruit(2002)]{spruit02}
 Spruit, H. C. 2002, \aap, 381, 923

\bibitem[Takeda(2000)]{takeda00}
 Takeda, Y. 2000, \pasj, 52, 113

\bibitem[Ud-Doula et al.(2009)]{ud-Doula09}
 Ud-Doula, A., Owocki, S. P., \& Townsend, R. H. D. 2009, \mnras, 392, 1022

\bibitem[Venn(1995)]{venn95}
 Venn, K. A. 1995, \apj, 449, 839

\bibitem[Venn \& Przybilla(2003)]{vepr03}
 Venn, K. A., \& Przybilla, N. 2003, in ASP Conf. Ser. 304, 20

\bibitem[Walborn(1976)]{walborn76}
 Walborn, N. R. 1976, \apj, 205, 419
 
\end{thebibliography}
\end{document}